\documentstyle[12pt,epsfig]{article}

\begin{document}
\begin {center}
{\bf {\Large
Transparency of the $\gamma (n,p) \pi^-$ reaction in nuclei
} }
\end {center}
\begin {center}
Swapan Das \\
{\it Nuclear Physics Division,
Bhabha Atomic Research Centre,  \\
Trombay, Mumbai 400085, India \\
Homi Bhabha National Institute, Anushakti Nagar,
Mumbai 400094, India }
\end {center}

\begin {abstract}
The transparency of the hadrons produced in the $\gamma (n,p) \pi^-$ reaction in nuclei is calculated using the 
Glauber model modified by including the Fermi motion of the nucleon in the nucleus. Since the calculated 
results underestimates the measured transparency for $^4$He nucleus, the Glauber model is further modified by 
incorporating the short-range correlation of the nucleon and the color transparency of the hadron in the 
nucleus. The nuclear transparency of the $\gamma (n,p) \pi^-$ reaction is calculated for $\theta_{\pi^-}$(c.m.) = 
50$^{\circ}$, 70$^{\circ}$ and 90$^{\circ}$. The calculated results are compared with the data reported for $^4$He 
nucleus.
\end {abstract}



\section{Introduction}

The nuclear transparency $T_A$ of a hadron $h$ can be defined by the ratio of the hadron-nucleus total cross 
section $\sigma_t^{hA}$  to the mass number $A$ (of the nucleus) times the hadron-nucleon total cross section 
$\sigma_t^{hN}$ in the free space. The interaction of the hadron with the nucleus reduces $\sigma_t^{hA}$ 
compared to $A \sigma_t^{hN}$, i.e., $T_A (=\frac{\sigma_t^{hA}}{A \sigma_t^{hN}}) < 1$. 
According 
to the Glauber model \cite{Glauber}, the hadron-nucleus cross section arises because of the multiple scattering 
or interaction of the hadron with nucleons in the nucleus. Therefore, the hadron-nucleon total cross section in 
the nucleus (denoted by $\sigma_t^{*hN}$) can be probed by studying $T_A$ of the hadron. 
The 
measured $T_A$ for the $\omega$ and $\phi$ mesons produced in the photonuclear reaction \cite{Wood} show both 
$\sigma_t^{*\omega N}$ and $\sigma_t^{*\phi N}$ are larger than the respective free space values, i.e., 
$\sigma_t^{*\omega N}$ $>$ $\sigma_t^{\omega N}$ and $\sigma_t^{*\phi N}$ $>$ $\sigma_t^{\phi N}$ \cite{DasA}.

One of the publicized issue in the nuclear physics is to search the color transparency of the hadron produced 
in the nucleus. The transverse size $d_\perp$ of a hadron produced in the nucleus due to the space-like 
four-momentum transfer $\Delta^2$ is shrinked as $d_\perp \approx 1/\Delta$ \cite{Farrar, Howell}. The reduced 
hadron (in size) is referred as the point like configuration (PLC) \cite{Howell}. 
The 
color neutral PLC, according to Quantum Chromodynamics, has reduced interaction with the nucleon in a nucleus 
\cite{Howell, DuttaH}. The PLC expands to the physical size of the hadron as it moves up to a length 
($\approx 1$ fm), called hadron formation length $l_h$ \cite{Howell}. The interaction of the PLC with the nucleon 
in a nucleus increases, as the size of the PLC enlarges during its passage up to $l_h$. The decrease of 
$\sigma_t^{*hN}$ compared to $\sigma_t^{hN}$ leads to the enhancement in $\sigma_t^{hA}$ \cite{Glauber}. As a 
result, $T_A$ of the hadron increases. This phenomenon is referred as color transparency (CT) of a hadron. Since 
$\Delta^2$ is involved, the CT is a energy dependent phenomenon. The physics of CT for hadrons have been discussed 
elaborately in Refs.~\cite{DuttaH, Frankfurt}.

It should be mentioned that $\sigma_t^{hA}$ can also be altered without the modification of $\sigma_t^{hN}$ in 
the nucleus. As explained by the Glauber model \cite{Glauber}, $\sigma_t^{hA}$ increases because of the short-range 
correlation (SRC) of the nucleon in the nucleus. 
The 
SRC originates because of the repulsive (short-range) interaction between the nucleons in the nucleus. The repulsive 
interaction keeps the bound nucleons apart ($\approx$ 1 fm), which is called nuclear granularity \cite{Lee}. 
Therefore, 
the SRC prevents the shadowing of the hadron-nucleon interaction due to the surrounding nucleons present in the 
nucleus. This phenomenon causes the enhancement in $\sigma_t^{hA}$, i.e., $T_A$ increases due to SRC of the nucleon. 
Unlike the color transparency, the SRC is independent of the energy-momentum of the hadron propagating through the 
nucleus.

The color transparency of the proton ($p$CT) is not found both in the A$(p,pp)$ and A$(e,e^\prime p)$ reactions. 
The nuclear transparency for protons in the previous reaction was measured at Brookhaven National Laboratory (BNL) 
\cite{Carroll}. The measured spectra could not be reproduced by the results calculated using the $p$CT in the 
Glauber model \cite{Lee}. The data can be understood by other mechanisms for the $pp$ scattering in the nucleus
(see Brodsky et al., \cite{Brodsky} and Ralston et al., \cite{Ralston}).
The 
proton transparency in the A$(e,e^\prime p)$ reaction was measured at Stanford Linear Accelerator Center 
(SLAC) \cite{Neill} and Jefferson Laboratory (JLab) \cite{DuttaW} for the photon virtuality $Q^2 = 0.64-14.2$ 
GeV$^2$. The data is well described by the short-range correlation of the bound nucleon included in the Glauber
model calculation \cite{DasTp}. The data are also analyzed for $Q^2<10$ GeV$^2$ by the other calculations which 
do not consider $p$CT \cite{Frankel, Lava}.

The data for the nuclear transparency of the meson is realized by the inclusion of the meson color transparency 
($m$CT) in the calculations. The $\rho$CT is reported in the experiment of the $\rho^{\circ}$ meson electroproduction 
from nuclei \cite{Airapetian}. There exist calculated results for the $\rho$CT in the energy region available at 
JLab \cite{Kopeliovich}. The nuclear transparency of the $K^+$ meson in the A$(e,e^\prime K^+)$ reaction evaluated 
using the $K$CT in the Glauber model calculation \cite{DasTK} is well accord with the the data reported from Jlab 
for $Q^2= 1.1-3.0$ GeV$^2$ \cite{Nuru}. 
The 
color transparency is also found in the nuclear diffractive dissociation of the pion (of 500 GeV/$c$) to dijets 
at Fermi National Accelerator Laboratory (FNAL) \cite{Aitala}. The $\pi$CT in the $(\pi^-,l^+l^-)$ reaction on 
nuclei is estimated  for $p_\pi = 5-20$ GeV/$c$ \cite{Larionov}, which can be measured at the forth-coming 
facilities in Japan Proton Accelerator Research Complex (J-PARC) \cite{Kumano}. This reaction provides 
informations complementary to those obtained from the A$(\gamma^*,\pi)$ reaction, i.e., A$(e,e^\prime \pi)$ 
reaction \cite{Clasie}.

The nuclear transparency of the $\pi^+$ meson produced in the $(e,e^\prime)$ reaction on nuclei was measured at 
JLab for $Q^2= 1.1-4.7$ GeV$^2$ \cite{Clasie}. The data have been understood due to the inclusion of the $\pi$CT 
in the Glauber model calculation \cite{DasTp, Larionov}. The $\pi$CT in the electronuclear reaction is also 
studied by Cosyn et al., \cite{Cosyn} and Kaskulov et al., \cite{Kaskulov} which reproduced the data \cite{Clasie} 
in the energy region available at JLab.
The
quoted transparency has been measured to explore the $\pi$CT in the higher $Q^2$ region, i.e., up to $\approx 10$ 
GeV$^2$ \cite{DuttaH, DuttaS}. The data will be reported in future. The calculated results for it can be seen in 
Ref.~\cite{DasTp}. Considering the $\pi$CT in this reaction, the dependence of the pion transparency on the 
momentum of the pion is also studied \cite{Larson}.

The meson is a quark-antiquark bound state where as the baryon is a composite state of three quarks. Therefore,
it can be thought that the PLC formation of a two-quark system is more probable than that of a three-quark system. 
The transparency of the meson and baryon can be studied simultaneously in the nuclear reaction.
Miller
and Strikman \cite{MillerS}, considering both $\pi$CT and $p$CT in the Glauber transparencies for the hadrons, have 
shown large enhancement in the transparency of the A$(\pi,\pi p)$ reaction at the energy 200 GeV available at CERN 
COMPASS experiment. Jain et al. \cite{JainK} predict the oscillation in the color transparency in the $\pi^-\mbox{A}$
and $\gamma \mbox{A}$ reactions, as that mentioned by Ralston et al., in the A$(p,pp)$ reaction \cite{Ralston}.

Recent past, the nuclear transparency of the $\gamma (n,p) \pi^-$ and $\gamma (n,p) \rho$ reactions in $^4$He and 
$^{12}$C nuclei were measured at Jlab for large cm energy $\sqrt{s}$ \cite{DuttaS}. The data of those reactions are 
not yet published. The measured transparency of the $\gamma (n,p) \pi^-$ process in $^4$He at low energy, i.e., 
$\sqrt{s}=1.99-2.95$ GeV, is reported from Jlab \cite{DuttaX}. The data are plotted versus the four-momentum transfer
$0.79-3.5$ GeV$^2$, which do not show the oscillation in the transparency. 
The 
nuclear transparency of the $\gamma (n,p) \pi^-$ reaction in $^4$He and $^{12}$C nuclei is investigated using the 
Glauber model. The nuclear phenomena, i.e., Fermi motion and short-range correlation of the nucleon, are used to 
modify this model. The color transparency of the hadron is also incorporated in the modified Glauber model.

\section{Formalism}

The hadron produced in the nuclear reaction interacts with the bound nucleons while propagating through the 
nucleus. This process at high energy can be described by the Glauber model \cite{Glauber}. Using this model,
the differential cross section of the elementary $\gamma (n,p)\pi^-$ reaction in a nucleus \cite{ParyevE}  
can be written as
\begin{equation}
\frac{d\sigma (\gamma A) }{ d\Delta^2 } = 
\int d{\bf r} \varrho_n ({\bf r}) 
 P_{\pi^-}({\bf r}) P_p({\bf r})
\left <  \frac{d\sigma_{\gamma n \to \pi^- p}}{d\Delta^2} (\sqrt s) \right >,
\label{dxA01}
\end{equation}
where $\Delta^2$ is the space-like four-momentum transfer in the elementary reaction. $\varrho_n ({\bf r})$ is 
the density of the neutron in the nucleus, normalized to the number of neutrons in the  nucleus. $P_h({\bf r})$ 
denotes the Glauber transparency of the hadron $h$ (i.e., the survival probability of $h$) when it traverses 
through the nucleus.

The quantity $\left <  \frac{d\sigma_{\gamma n \to \pi^- p}}{d\Delta^2} (\sqrt s) \right >$ represents the 
differential cross section of the elementary $\gamma n \to \pi^- p$ reaction in the nucleus. Since the bound 
nucleon possesses Fermi motion, it can be expressed as \cite{ParyevE}:
\begin{equation}
\left <  \frac{d\sigma_{\gamma n \to \pi^- p}}{d\Delta^2} (\sqrt s) \right > 
= \int \int d{\bf k}_i d\epsilon_i P(k_i,\epsilon_i)  
                                   \frac{d\sigma_{\gamma n \to \pi^- p}}{d\Delta^2} (\sqrt s),
\label{dxN01}
\end{equation}
where $P(k_i,\epsilon_i)$ denotes the spectral function of the target nucleus, normalized to unity. It 
represents the probability of finding a nucleon of momentum ${\bf k}_i$ and binding energy $\epsilon_i$ in 
the  nucleus \cite{ParyevE, Benhar}. $P(k_i,\epsilon_i)$ has been discussed elaborately in 
Ref.~\cite{ParyevK}.
$\frac{d\sigma_{\gamma n \to \pi^- p}}{d\Delta^2} (\sqrt s)$ is the cross section of the 
$\gamma n \to \pi^- p$ reaction in the free space occurring at the c.m. energy  
$\sqrt s = \sqrt { (E_\gamma +E_N)^2 + ({\bf k}_\gamma +{\bf k}_i)^2 }$. The energy of the nucleon  in the 
nucleus is $E_N = m_A - \sqrt { (-{\bf k}_i)^2 + (m_A-m_N+\epsilon_i)^2 }$, where $m_N$ and $m_A$ are the 
masses of the nucleon (in the free state) and nucleus respectively.

The survival probability of the hadron, i.e., $P_h({\bf r})$ in Eq.~(\ref{dxA01}), propagating through the 
uncorrelated nucleons in the nucleus \cite{Gao} is given by
\begin{equation}
P_h({\bf r}) = 
exp \left \{ -\int_0^\infty dl \varrho ({\bf r} + {\bf {\hat k}}_h l) \sigma_t^{hN} \right \},
\label{Ph01}
\end{equation}
where $\sigma_t^{hN}$ is the hadron-nucleon total cross section in the free space. $l$ is the distance in the 
nucleus travelled by the hadron $h$ in the direction of its momentum ${\bf \hat {k}}_h$. $\varrho$ is the 
density distribution, normalized to the mass number of the nucleus.

$P_h({\bf r})$ increases because of the correlated nucleon (i.e., short-range correlation (SRC) of the nucleon) 
in the nucleus. Therefore, the hadron-nucleus cross section (as discussed earlier) increases due to the SRC 
of the nucleon. This occurs since the SRC of the nucleon in the nucleus modifies its density distribution 
$\varrho$ in Eq.~(\ref{Ph01}) \cite{Lee} as  
\begin{equation}
\varrho ({\bf r} + {\bf {\hat k}}_h l) \to \varrho ({\bf r} + {\bf {\hat k}}_h l) C(l),
\label{rhcr}
\end{equation}
where $C(l)$ represents the correlation function. Using the nuclear matter estimate, it is given by
\begin{equation}
C(l) = \left [ 1-\frac{h(l)^2}{4} \right ]^{1/2} [1+f(l)],
\label{crfn}
\end{equation}
with $h(l)=3\frac{j_1(k_F l)}{k_F l}$ and $f(l)=-e^{-\alpha l^2} (1-\beta l^2)$. The  Fermi momentum $k_F$ is 
chosen equal to 1.36 fm$^{-1}$. $C(l)$ with the parameters $\alpha =1.1$ fm$^{-2}$ and $\beta =0.68$ fm$^{-2}$ 
agrees well that derived from the  many-body calculations \cite{Lee}.

The hadron-nucleon cross section in a nucleus, as mentioned  earlier, can be less than that in the free space 
due to the color transparency of the hadron ($h$CT). The reduction in the cross section causes the enhancement 
in $P_h({\bf r})$ in Eq.~(\ref{Ph01}), and hence the nuclear transparency of the hadron increases. To look for 
the $h$CT, $\sigma^{hN}_t$ in $P_h({\bf r})$ (according to quantum diffusion model \cite{Farrar, Howell}) has to 
replace by that in the nucleus, i.e., $\sigma^{*hN}_t$:
\begin{equation}
\sigma^{*hN}_t =
\sigma^{hN}_{t,CT} (\Delta^2, l) =\sigma^{hN}_t
\left [
\left \{
\frac{l}{l_h} +\frac{n_q^2<k^2_t>}{\Delta^2} \left ( 1 -\frac{l}{l_h} \right )
\right \} \theta(l_h-l)
+ \theta(l-l_h) \right ],
\label{XCT}
\end{equation}
where 
$\sigma^{hN}_{t,CT} (\Delta^2, l)$ refers $\sigma^{*hN}_t$ due to the $h$CT. $n_q$ denotes the number of the
valence quarks or quak-antiquark present in the hadron, e.g., $n_q=2(3)$ for pion  (proton) \cite{Farrar}. 
$<k_t^2>^{1/2} (=0.35$ GeV/$c$) illustrates the transverse momentum of the (anti)quark. $l$ is that defined 
in Eq.~(\ref{Ph01}). As illustrated earlier, $l_h$ represents the hadron formation-length which is expressed 
\cite{Larionov} as
\begin{equation}
l_h = \frac{2k_h}{\Delta M^2},
\label{Lh}
\end{equation}
where $k_h$ is the momentum of the hadron in the laboratory frame. $\Delta M^2$ is related to the mass 
difference of the hadronic states originating due to the fluctuation of the (anti)quark in the PLC of the 
hadron \cite{Farrar, Howell}.

\section{Result and Discussions}

The transparency of the hadrons produced in the $\gamma (n,p) \pi^-$ process in nuclei has been calculated. 
The data for the $^4$He nucleus are reported by Dutta et al. \cite{DuttaX} in the four-momentum transfer 
region $0.79 - 3.5$ GeV$^2$. They have extracted transparency from the measured and Monte Carlo yields from 
$^4$He and $d$ (deuteron) nuclei, using the relation 
\cite{DuttaX}
\begin{equation}
T(^4He)
= \frac{ \frac{ Yeild_{Data} (^4He) }{ Yeild_{Monte Carlo} (^4He) } }
       { \frac{ Yeild_{Data} (d)  }{ Yeild_{Monte Carlo} (d)  } } T(d).
\label{trex}
\end{equation}
The ratio removes the uncertainties in the results. The transparency for $d$, i.e., $T(d)$, is obtained from 
the measured proton transparency in the $d(e,e^\prime p$) reaction and the $\pi^-$ meson transparency in the 
deuteron nucleus. The value of $T(d)$ is around 0.8, as tabulated in  Ref.~\cite{DuttaX}.

The cross section of the $\gamma (n,p) \pi^-$ reaction in the nucleus is calculated using Eq.~(\ref{dxA01}) 
to estimate the transparency of the reaction. Following Eq.~(\ref{trex}), the quoted transparency is 
calculated as 
\begin{equation}
T(A) = 
\frac{
\frac{ d\sigma (\gamma A) / d\Delta^2 }{ d\sigma(\gamma A) / d\Delta^2_{PWIA} }
      }{
\frac{ d\sigma (\gamma d) / d\Delta^2 }{ d\sigma(\gamma d) / d\Delta^2_{PWIA} }
      }
T(d).
\label{trcl}
\end{equation}
The suffix $PWIA$ stands for the plane wave impulse approximation where the final state interactions of the 
hadrons are neglected.

The differential cross sections for the four-momentum transfer distribution of the $\gamma (n,p) \pi^-$ reaction are 
measured by Zhu et al., \cite{Zhu} at $\theta_{\pi^-} ~(\mbox{c.m.})$ = 50$^{\circ}$, 70$^{\circ}$ and 90$^{\circ}$.
Those cross sections are used in Eq.~(\ref{dxN01}) to calculate the transparency of the $\gamma (n,p) \pi^-$ 
reaction in the nucleus. The measured total cross sections $\sigma_t^{\pi^- N}$ and $\sigma_t^{pN}$ \cite{Zyla} 
are used to evaluate the survival probabilities [in Eq.~(\ref {Ph01})] for the pion and proton respectively. 
The density distribution of deuteron is generated using Hulthen wave function \cite{Adler}, where as that for 
other nuclei (as reported from the electron scattering experiment) is taken from Ref.~\cite{Jager}.

As described earlier, the nuclear transparency of the proton measured in the A$(e,e^\prime p)$ reaction for 
the wide range of the four-momentum transfer \cite{Neill, DuttaW} do not show the color transparency (CT) of 
the proton. The data is reproduced by considering the short-range correlation (SRC) of the nucleon, as 
described in Eq.~(\ref{crfn}), in the Glauber model calculation \cite{DasTp} (also see the references 
therein). Therefore, the SRC (but not the CT) is included in the Glauber model to estimate the survival 
probability of the proton, i.e., $P_p ({\bf r})$ in Eq.~(\ref{Ph01}).
The 
measured pionic transparency in the A$(e,e^\prime \pi^+)$ reaction \cite{Clasie} is reproduced well due to the 
incorporation of the pion color transparency $\pi$CT, as illustrated in Eq.~(\ref{XCT}), in the Glauber model 
calculation \cite{DasTp, Larionov}. Therefore, the $\pi$CT in the Glauber model is used to evaluate the 
survival probability of the pion, see $P_{\pi^-} ({\bf r})$ in Eq.~(\ref{Ph01}).

\begin{figure}[h]
\begin{center}
\centerline {\vbox {
\psfig{figure=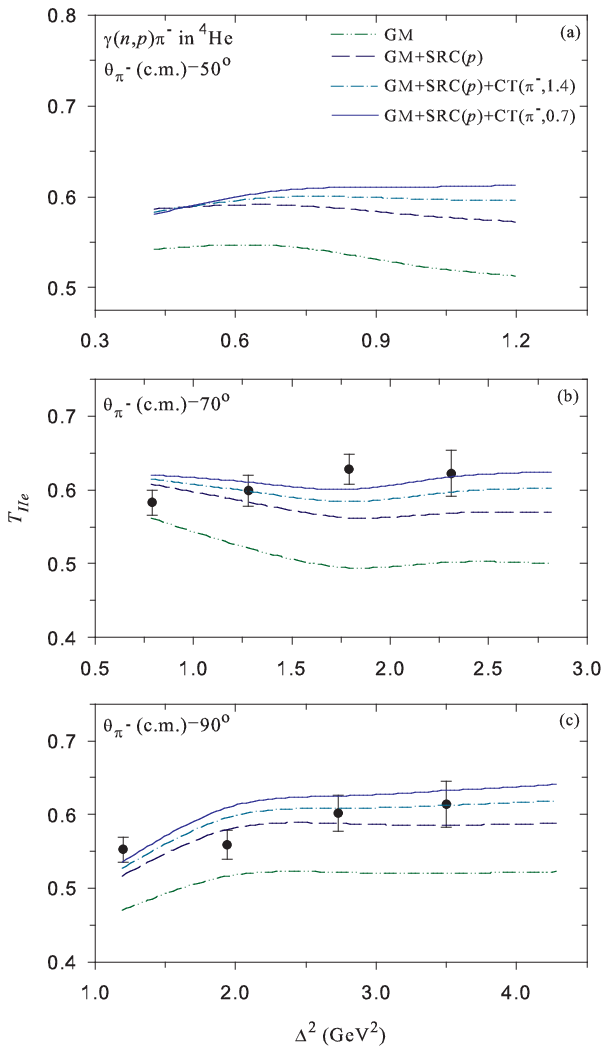,height=18.0 cm, width=08.0 cm}
}}
\caption{
(color online).
The transparency $T_{He}$ of the pion and proton in $^4$He vs the four-momentum transfer $\Delta^2$. The data 
are taken from Ref.~\cite{DuttaX}. The curves appearing in the figure have been explained in the text.
}
\label{FTAHe}
\end{center}
\end{figure}

The calculated nuclear transparency for the $\gamma (n,p) \pi^-$ reaction in the $^4$He nucleus, denoted by $T_{He}$, 
is shown in Fig.~\ref{FTAHe} along with the data for $\theta_{\pi^-}$ (c.m.) = 70$^{\circ}$ and 90$^{\circ}$ 
\cite{DuttaX}. The dot-dot-dashed curve [labelled as GM] arises due to the Glauber model (GM) calculation. The 
long-dashed curve [labelled as GM+SRC($p$)] describes the transparency evaluated using the short-range 
correlation (SRC) of the bound nucleon in the Glauber model calculation for the proton survival probability 
$P_p ({\bf r})$ in Eq.~(\ref{Ph01}). 
The 
dot-dashed curve [labelled as GM+SRC($p$)+CT($\pi^-,1.4$)] represents the calculated results, where the SRC 
is incorporated in the Glauber model to estimate $P_p ({\bf r})$ and $\pi$CT with $\Delta M^2 =1.4$ GeV$^2$ 
[defined in Eq.~(\ref{Lh})] is included in the Glauber model to evaluate $P_{\pi^-} ({\bf r})$ in 
Eq.~(\ref{Ph01}). The solid curve [labelled as GM+SRC($p$)+CT($\pi^-,0.7$)] illustrates that of the dot-dashed 
curve except $\Delta M^2$ taken equal to 0.7 GeV$^2$.

\begin{figure}[h]
\begin{center}
\centerline {\vbox {
\psfig{figure=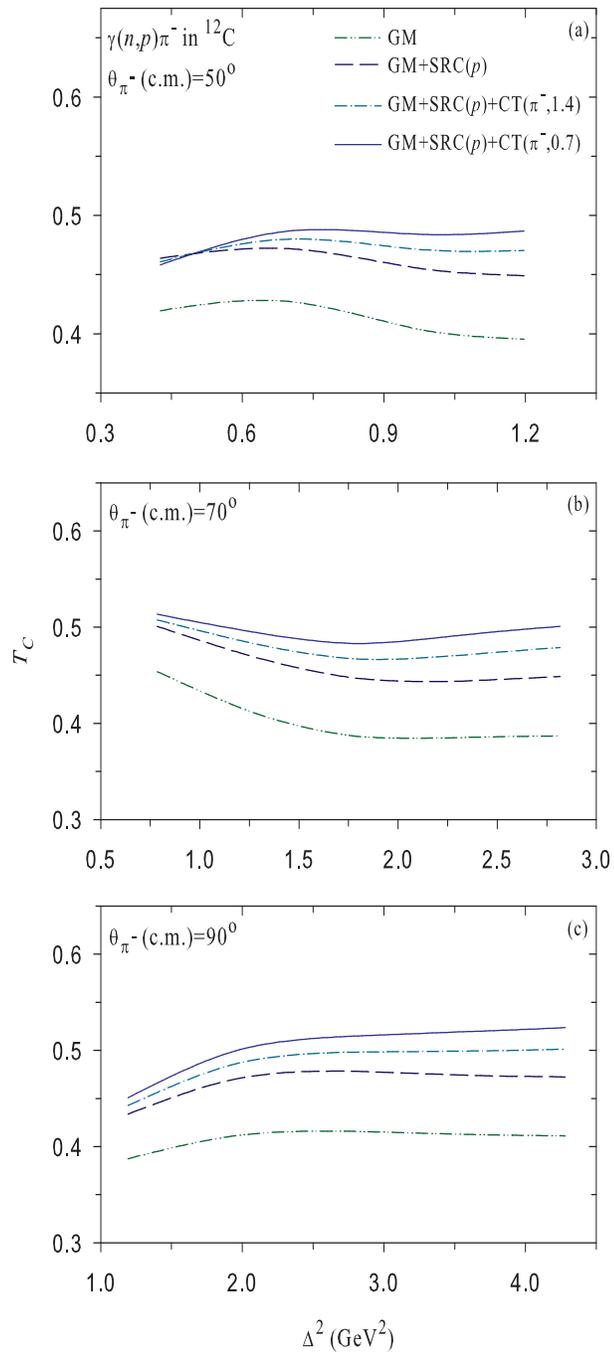,height=18.0 cm, width=08.0 cm}
}}
\caption{
(color online).
Same as those described in Fig.~\ref{FTAHe}, but the transparency $T_C$ presented for $^{12}$C nucleus.
}
\label{FTAC}
\end{center}
\end{figure}

The transparency of the $\gamma (n,p) \pi^-$ reaction in $^{12}$C nucleus, denoted by $T_C$, have been calculated 
using the (modified) Glauber model, as done for $^4$He nucleus. As shown in Fig.~\ref{FTAC}, the calculated $T_C$ 
are qualititavely similar to $T_{He}$ presented in Fig.~\ref{FTAHe}. The magnitude of $T_{C}$ is less than that of 
$T_{He}$, since the survival probability reduces for the hadrons propagating through the large nucleus.

\section{Conclusions}

The nuclear transparency of the hadrons produced in the $\gamma (n,p) \pi^-$ reaction has been calculted for 
$^4$He and $^{12}$C nuclei using the Glauber model where the Fermi motion of the nucleon in the nucleus is 
incorporated. The calculated results underestimate the data reported for the $^4$He nucleus. Therefore, the 
Glauber model is further modified by including the short-range correlation of the nucleon to evaluate the 
proton survival probability and the color transparency to determine the pion survival probability. Those 
modifications are done based on the ealier studies.
The 
calculated results are compared with the data reported for the $^4$He nucleus in the four-momentum transfer 
region $0.79-3.5$ GeV$^2$, which shows more data are required to conclude the color transparency of the pion. 
The ongoing analysis of the experimental results for $^4$He and $^{12}$C nuclei for the wide range of the 
four-momentum transfer can highlight on this issue in future.

\section{Acknowledgement}

The author thanks the referee for the comment which helps to improve the quality of the paper. Dipangkar Dutta is 
gratefully acknowledged for the discussions on the experimental results. The author is thankful to A. K. Gupta and 
S. M. Yusuf for the encouragement to work on theoretical nuclear physics.





\newpage

\newpage

\end{document}